\documentclass[journal]{IEEEtran}

\usepackage{amsmath}
\usepackage{amssymb}
\usepackage{amsthm}
\usepackage{xcolor}
\usepackage{graphicx}
\usepackage{amsmath}
\usepackage{amsfonts}
\usepackage{url}
\usepackage[compress]{cite}
\usepackage{tikz}

\newcommand{\range}[1]{[#1]}
\DeclareMathOperator{\supp}{supp}
\DeclareMathOperator{\trace}{trace}

\newtheorem{assumption}{Assumption}
\newtheorem{theorem}{Theorem}
\newtheorem{corollary}{Corollary}

\begin{document}

\title{A Fundamental Bound on Performance of Non-Intrusive Load Monitoring \\ with Application to Smart Meter Privacy}

\author{Farhad Farokhi\thanks{F. Farokhi is with the CSIRO's Data61 and the University of Melbourne. The work of F. Farokhi was, in part, funded by the office of the Deputy Vice-Chancellor (Research) at the University of Melbourne.  e-mail: farhad.farokhi@unimelb.edu.au}} 

\maketitle

\begin{abstract}   We prove that the expected estimation error of non-intrusive load monitoring algorithms is lower bounded by the trace of the inverse of the cross-correlation matrix between the derivatives of the load profiles of the appliances. We use this  fundamental bound to develop privacy-preserving policies. Particularly, we devise a load-scheduling policy by maximizing the lower bound on the expected estimation error of non-intrusive load monitoring algorithms.
\end{abstract}


\section{Introduction}
Non-intrusive load monitoring research is dedicated to development of algorithms for dis-aggregating overall energy consumption of households measured by smart meters to estimate timing of individual appliances, such as fridge or air-conditioning units~\cite{hosseini2017non, zoha2012non,davies2019deep, jin2011robust,zhao2015blind, rashid2019evaluation}. The research is often motivated by the interest in providing consumers with energy-saving tips to lower bills or counter climate change, performing fault detection, and accommodating electricity grid transformations due to integration of renewable energy. Although an important area of research, little is done in understanding fundamental bounds on the achievable performance of non-intrusive load monitoring algorithms. Therefore, in this paper, we prove that \textit{the expected estimation error of non-intrusive load monitoring algorithms is lower bounded by the trace of the inverse of the cross-correlation matrix between the derivatives of the load profiles of the appliances}.   

Energy data, collected by smart meters, is known to leak private information of households, such as occupancy and appliance usage~\cite{mcdaniel2009security}. This sensitive data can be accessed by third-party data-analytic companies\footnote{\label{note1} See \url{http://bidgely.com} and \url{https://plotwatt.com/} as examples.} through electricity retailers or utility companies. For instance, in Australia, an electricity retailer required its customers to consent to sharing their data with third parties in the United States before permitting them to use an online web portal~\cite{ChadwickButtCook2012}.  Therefore, non-intrusive load monitoring can be used for gaining privacy-intrusive insights. This is evident from the patents on the technology expressing its use in targeted advertising~\cite{leeb1996transient, haghighat2015system}. Also, non-intrusive load monitoring has been proved to be commercially viable and attractive. Similar technologies are currently being used (albeit based on water smart meters) to track elderly behaviour and help those in distress~\cite{CornerAussie2018} with similar applications to monitoring patients~\cite{chalmers2018intelligent}, which is a double-edged sword that can be used by both healthcare professionals and  insurance agencies. This has motivated the development of privacy-preserving polices for smart meters; see~\cite{yao2017privacy, cho2019effect, liu2017information, sankar2012smart, eibl2014influence, farokhi2017fisher, giaconi2017smart, liu2017achieving} and references there-in. Therefore, in this paper, we use the aforementioned  \textit{fundamental bound on estimation error of non-intrusive load monitoring algorithms to develop privacy-preserving policies}. Particularly, we devise a load-scheduling policy by maximizing the trace of the inverse of the cross-correlation matrix between the derivatives of the load profiles of the appliances.

\section{Non-Intrusive Load Monitoring}
Consider a house with $n\in\mathbb{N}$ appliances. Appliance $i\in\range{n}:=\{1,\dots,n\}$ has a load signature of $f_i:\mathbb{R}\rightarrow\mathbb{R}$, i.e., its unique energy consumption pattern. The load's signature is such that $f_i(t)=0$ for all $t<0$ (before starting to work) and $t>T$ (after finishing its work) for some large enough $T$. Appliance $i\in\range{n}$ is scheduled to start at $\tau_i\in\mathbb{R}$. Therefore, the total consumption of the house at any given time $t\in\mathbb{R}$ is $\sum_{i\in\range{n}} f_i(t-\tau_i)$. 

We assume that a non-intrusive load monitoring algorithm can access noisy measurements of the total consumption at discrete times  $(t_\ell)_{\ell\in\range{k}}\subseteq [0,T]$. The noise models measurement noise, privacy-preserving additive noises (e.g., differential-privacy noise), and consumption of small loads that the non-intrusive load monitoring algorithm is not interested in identifying. The measurement $y_\ell$ at time $t_\ell$ is then given by
\begin{align}
y_\ell=\sum_{i\in\range{n}} f_i(t_\ell-\tau_i)+w_\ell,
\end{align}
where $w:=(w_\ell)_{\ell \in\range{k}}$ is a sequence of i.i.d.\footnote{i.i.d. stands for independently and identically distributed.} noises. The non-intrusive load monitoring algorithm is interested in estimating $\tau:=(\tau_i)_{i\in\range{n}}\in\mathbb{R}^n$ from the measurements $y:=(y_\ell)_{\ell\in\range{k}}\in\mathbb{R}^k$ using a family of arbitrary estimators denoted by $\hat{\tau}_i:\mathbb{R}^k\rightarrow\mathbb{R}$ for all $i\in\range{n}$. We are interested for finding a lower bound on 
\begin{align}
&\sum_{i\in\range{n}}\pi_i\mathbb{E}\{(\hat{\tau}_i(y)-\tau_i)^2\}
=\mathbb{E}\{\|\Pi^{1/2}(\hat{\tau}(y)-\tau)\|_2^2\},
\end{align}
where $\hat{\tau}:= (\hat{\tau}_i)_{i\in\range{n}}$ and $\Pi:=\mathrm{diag}(\pi_1,\dots,\pi_n)$. We make the following standing assumption.

\begin{assumption}[Regularity] \label{assum:1} $p(w)$ is continuously differentiable; $p(w)=0, \forall w\in\partial \supp(p)$.
\end{assumption}

The regularity condition in Assumption~\ref{assum:1} is a basic assumption that is common in signal processing results, such as the Cram\'{e}r-Rao bound~\cite[p.\,169]{cramerraotheorem}.  This assumption holds for Gaussian and Laplace distributions, and many other density functions. In fact, any density function with an unbounded support automatically satisfies this condition. We can prove the following fundamental bound on the performance of unbiased non-intrusive load monitoring algorithms.

\begin{theorem} \label{tho:1} For any unbiased estimator $\hat{\tau}$, i.e.,  $\mathbb{E}\{\hat{\tau}\}=\tau$, we get
\begin{align*}
\mathbb{E}\{\|\Pi^{1/2}(\hat{\tau}(y)-\tau)\|_2^2\}\geq \frac{1}{\mathcal{I}^w}\trace(\Pi R_{\mathrm{d}}(\tau)^{-1}),
\end{align*}
where $\mathcal{I}^w=\mathbb{E}_w\{(p'(w)/p(w))^2\}$ is the Fisher information of the additive noise and $R_{\mathrm{d}}(\tau)$ is the discrete cross-correlation matrix function for the derivatives of the load signatures with entry in $i$-the row and $j$-the column defined as
\begin{align*}
[R_{\mathrm{d}}]_{ij}(\tau_i,\tau_j):=\sum_{\ell\in\range{k}}f'_i(t_\ell-\tau_i)f'_j(t_\ell-\tau_j).
\end{align*}
\end{theorem}

\begin{IEEEproof} The conditional probability density of observing $y$ given $\tau$ is equal to
\begin{align*}
p(y|\tau)=\prod_{\ell\in\range{k}} p\left( y_\ell-\sum_{i\in\range{n}} f_i(t_\ell-\tau_i)\right).
\end{align*}
Differentiating logarithm of the conditional density $p(y|\tau)$ results in
\begin{align*}
\frac{\partial \log( p(y|\tau))}{\partial \tau_i} =\hspace{-.03in}\sum_{j\in\range{k}} \frac{\displaystyle p'\hspace{-.03in}\left(\hspace{-.03in} y_j-\hspace{-.03in}\sum_{i\in\range{n}} \hspace{-.03in}f_i(t_j-\tau_i)\hspace{-.03in}\right)}{\displaystyle p\hspace{-.03in}\left( \hspace{-.03in}y_j-\hspace{-.03in}\sum_{i\in\range{n}} \hspace{-.03in}f_i(t_j-\tau_i)\hspace{-.03in}\right)} f'_i(t_j-\tau_i).
\end{align*}
Following this, we can compute the entry in $i$-th row and $q$-th column of the Fisher information matrix as
\begin{align*}
\mathcal{I}_{iq}:=
&\mathbb{E}_{y}\left\{\frac{\partial}{\partial \tau_i}	\log( p(y|\tau))\frac{\partial}{\partial \tau_q}	\log( p(y|\tau))\right\}\\
=&\mathbb{E}_{w}\Bigg\{\Bigg( \sum_{j\in\range{k}} \frac{\displaystyle p'(w_j)}{\displaystyle p(w_j)} f'_i(t_j-\tau_i)\Bigg)\\
&\hspace{.5in}\times \Bigg( \sum_{j\in\range{k}} \frac{\displaystyle p'(w_j)}{\displaystyle p(w_j)} f'_q(t_j-\tau_q)\Bigg)\Bigg\}\\
=&\sum_{j_1,j_2\in\range{k}} \mathbb{E}_{w}\Bigg\{\frac{\displaystyle p'(w_{j_1})}{\displaystyle p(w_{j_1})} \frac{\displaystyle p'(w_{j_2})}{\displaystyle p(w_{j_2})}\Bigg\} \\
&\hspace{.5in}\times  f'_i(t_{j_1}-\tau_i)f'_q(t_{j_2}-\tau_q)\\
=& \mathcal{I}^w \sum_{j\in\range{k}}f'_i(t_j-\tau_i)f'_q(t_j-\tau_q)\\
=& \mathcal{I}^w [R_{\mathrm{d}}]_{iq}(\tau_i,\tau_q).
\end{align*}
Therefore, the Fisher information matrix is equal to
\begin{align*}
\mathcal{I}:=&\mathbb{E}_{y}\left\{\nabla_\tau p(y|\tau)\nabla_\tau p(y|\tau)^\top  \right\}
=\mathcal{I}^w R_{\mathrm{d}}(\tau).
\end{align*}
Finally, we get
\begin{align*}
\mathbb{E}\{\|\Pi^{1/2}(\hat{\tau}(y)&-\tau )\|_2^2\}\\
=&\trace(\mathbb{E}\{\Pi^{1/2}(\hat{\tau}-\tau )(\hat{\tau}-\tau )^\top \Pi^{1/2} \})\\
=&\trace(\Pi^{1/2}\mathbb{E}\{(\hat{\tau}-\tau )(\hat{\tau}-\tau )^\top  \}\Pi^{1/2})\\
\geq &\trace(\Pi^{1/2}(\mathcal{I}^w R_{\mathrm{d}}(\tau))^{-1}\Pi^{1/2})\\
= &\trace(\Pi^{1/2} R_{\mathrm{d}}(\tau)^{-1}\Pi^{1/2})/\mathcal{I}^w\\
= &\trace(\Pi R_{\mathrm{d}}(\tau)^{-1})/\mathcal{I}^w.
\end{align*}
This concludes the proof.
\end{IEEEproof}

If we assume that $w_\ell$ is a zero-mean Gaussian random variable with variance $\sigma_w^2>0$ for all $\ell\in\range{k}$, we can simplify the bound in Theorem~\ref{tho:1} by noting that $\mathcal{I}^w=\sigma_w^{-2}$.

Note that the bound in Theorem~\ref{tho:1} is only valid for unbiased estimators. We generalize this bound to unbiased estimators in the next theorem.
%
%
%

\begin{theorem} \label{tho:2} For any estimator $\hat{\tau}$, we get
	\begin{align*}
	\mathbb{E}\{&\|\Pi^{1/2}(\hat{\tau}(y)-\tau)\|_2^2\}\\ \geq& \frac{1}{\mathcal{I}^w}\trace\Bigg(\Pi \frac{\partial \mu(\tau) }{\partial \tau}R_{\mathrm{d}}(\tau)^{-1}\frac{\partial \mu(\tau) }{\partial \tau}^\top\Bigg)+\|\Pi^{1/2}\mu(\tau)\|_2^2,
	\end{align*}
	where $\mu(\tau):=\mathbb{E}\{\hat{\tau}(y)\}$.
\end{theorem}

\begin{IEEEproof}
The proof follows from that
\begin{align*}
\mathbb{E}\{\|\Pi^{1/2}(\hat{\tau}&(y)-\tau )\|_2^2\}\\
\geq &\trace\Bigg(\Pi^{1/2}\Bigg(\frac{\partial \mu(\tau) }{\partial \tau}(\mathcal{I}^w R_{\mathrm{d}}(\tau))^{-1}\frac{\partial \mu(\tau) }{\partial \tau}^\top\\
&+\mu(\tau)\mu(\tau)^\top\Bigg)\Pi^{1/2}\Bigg)\\
= &\frac{1}{\mathcal{I}^w}\trace\Bigg(\Pi \frac{\partial \mu(\tau) }{\partial \tau}R_{\mathrm{d}}(\tau)^{-1}\frac{\partial \mu(\tau) }{\partial \tau}^\top\Bigg)\\
&+\|\Pi^{1/2}\mu(\tau)\|_2^2.
\end{align*}
This concludes the proof.
\end{IEEEproof}

We can further simplify the lower bound in Theorem~\ref{tho:2}.
 
 \begin{corollary} \label{cor:1}
For any estimator $\hat{\tau}$, we get
\begin{align*}
\mathbb{E}\{\|\Pi^{1/2}(\hat{\tau}(y)-\tau)\|_2^2\}\geq& \frac{c_1(\tau)}{\mathcal{I}^w}\trace(\Pi R_{\mathrm{d}}(\tau)^{-1}) +c_2(\tau),
\end{align*}
where 
\begin{align*}
c_1(\tau):=\lambda_{\min}\Bigg(\Pi^{-1/2}\frac{\partial \mu(\tau) }{\partial \tau}^\top\Pi \frac{\partial \mu(\tau) }{\partial \tau}\Pi^{-1/2}\Bigg)\geq 0,\\
c_2(\tau):=\|\Pi^{1/2}\mu(\tau)\|_2^2\geq 0.
\end{align*}
 \end{corollary}

\begin{IEEEproof}
Note that
\begin{align*}
\trace\Bigg(\Pi& \frac{\partial \mu(\tau) }{\partial \tau}R_{\mathrm{d}}(\tau)^{-1}\frac{\partial \mu(\tau) }{\partial \tau}^\top\Bigg)\\
&=\trace\Bigg(\Pi^{-1/2}\frac{\partial \mu(\tau) }{\partial \tau}^\top\Pi \frac{\partial \mu(\tau) }{\partial \tau}\Pi^{-1/2}\\
&\hspace{1.2in}\times\Pi^{1/2} R_{\mathrm{d}}(\tau)^{-1}\Pi^{1/2}\Bigg)\\
&\geq \lambda_{\min}\Bigg(\Pi^{-1/2}\frac{\partial \mu(\tau) }{\partial \tau}^\top\Pi \frac{\partial \mu(\tau) }{\partial \tau}\Pi^{-1/2}\Bigg)\\
&\hspace{1.2in}\times\trace(\Pi^{1/2} R_{\mathrm{d}}(\tau)^{-1}\Pi^{1/2}),
\end{align*}
where the inequality follows from the inequality on traces of positive semi-definite matrices  in~\cite{kleinman1968design}.
 Combining this inequality with $\trace(\Pi^{1/2} R_{\mathrm{d}}(\tau)^{-1}\Pi^{1/2})=\trace(\Pi R_{\mathrm{d}}(\tau)^{-1})$ concludes the proof.
\end{IEEEproof}

Note that $R_{\mathrm{d}}(\tau)$ in Theorems~\ref{tho:1} and~\ref{tho:2} is a function of the sampling times $(t_{\ell})_{\ell\in\range{k}}$. However, as $k$ increases, this dependence disappears. This is discussed in the following corollary.

\begin{corollary} \label{cor:2}
For any unbiased estimator $\hat{\tau}$, i.e.,  $\mathbb{E}\{\hat{\tau}\}=\tau$, we get
\begin{align*}
\lim_{k\rightarrow \infty} k\mathbb{E}\{\|\Pi^{1/2}(\hat{\tau}-\tau)\|_2^2\}\geq \frac{1}{\mathcal{I}^w}\trace(\Pi R_{\mathrm{c}}(\tau)^{-1}),
\end{align*}
where $R_{\mathrm{c}}(\tau)$ is the continuous cross-correlation matrix function for the derivatives of the load signatures with entry on $i$-the row and $j$-the column defined as
\begin{align*}
[R_{\mathrm{c}}]_{ij}(\tau_i,\tau_j):=\int_{-\infty}^{+\infty} f'_i(t-\tau_i)f'_j(t-\tau_j)\mathrm{d}t.
\end{align*}
\end{corollary}

\begin{IEEEproof}
The proof follows from the convergence of the Riemann integral. The bounds of the integral can be pushed from $[0,T]$ to $(-\infty,+\infty)$ due to the fact that the load profiles are equal to zero outside $[0,T]$.
\end{IEEEproof}

%
%
\begin{figure}
	\begin{tikzpicture}
	\node[] at (0,0) {\includegraphics[width=1\linewidth]{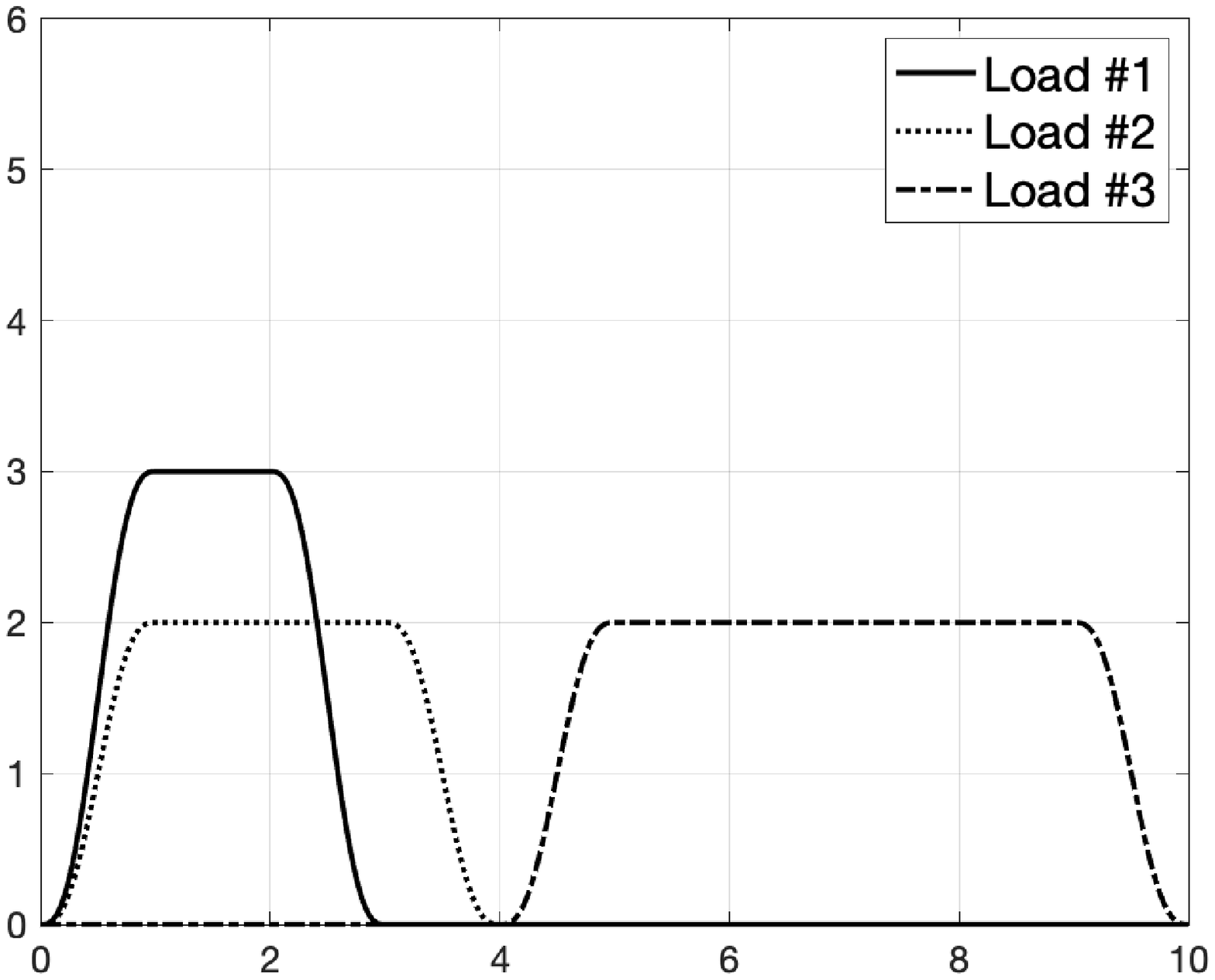}};
	\node[] at (0,-3.2) {$t$};
	\node[rotate=90] at (-4,0) {$f_i(t)$};
	\end{tikzpicture}
	\vspace{-.3in}
	\caption{
		\label{fig:example}
		Examples of three generic loads.}	
\end{figure}
\section{Smart -Meter Privacy by Load Scheduling}
In this section, we propose an approach for smart meter privacy based on load scheduling.
Clearly, the lower bound on the performance of an adversary employing a non-intrusive load monitoring algorithm for prying into a household in Theorem~\ref{tho:1} and Corollary~\ref{cor:1} is a function of the scheduling time of the appliances $\tau.$ This motivate us to solve the following optimization problem for finding the optimal privacy-preserving scheduling:
\begin{align} \label{eqn:optimi1}
\max_{\tau\in \mathfrak{T}} \trace(\Pi R_{\mathrm{d}}(\tau)^{-1}),
\end{align}
where $\mathfrak{T}$ denotes the set of feasible schedules. The cost function of this optimization problem is however non-convex. Therefore, there are many local optima that can be recovered using numerical algorithms, such as the gradient descent.  For this, we note that
\begin{align*}
\hspace{+.5in}&\hspace{-.5in}\frac{\partial \trace(\Pi R_{\mathrm{d}}(\tau)^{-1})}{\partial \tau_i}\\
&=\trace\Bigg(\Pi\Bigg(\frac{\partial R_{\mathrm{d}}(\tau)^{-1}}{\partial \tau_i}\Bigg)\Bigg)\\
&=-\trace\hspace{-.03in}\Bigg(\hspace{-.03in}R_{\mathrm{d}}(\tau)^{-1}\Pi R_{\mathrm{d}}(\tau)^{-1}\Bigg(\hspace{-.03in} \frac{\partial R_{\mathrm{d}}(\tau)}{\partial \tau_i} \hspace{-.03in}\Bigg)\hspace{-.03in}\Bigg),
\end{align*}
where
\begin{align*}
\bigg[\frac{\partial R_{\mathrm{d}}(\tau)}{\partial \tau_i}\bigg]_{ji}&=\bigg[\frac{\partial R_{\mathrm{d}}(\tau)}{\partial \tau_i}\bigg]_{ij}\\
&=\begin{cases}
-\displaystyle 2\sum_{j\in\range{k}}f''_i(t_j-\tau_i)f'_i(t_j-\tau_i), & i=j,\\
-\displaystyle \sum_{j\in\range{k}}f''_i(t_j-\tau_i)f'_j(t_j-\tau_j), & i\neq j,
\end{cases}
\end{align*}
and
\begin{align*}
\bigg[\frac{\partial R_{\mathrm{d}}(\tau)}{\partial \tau_i}\bigg]_{q\ell}=0,\quad \forall q,\ell\neq i.
\end{align*}
Therefore, we can use the projected gradient ascent to solve~\eqref{eqn:optimi1} by following
\begin{align}
\tau^{k+1}=P_{\mathfrak{T}} \left[ \tau^{k}+\mu_k\frac{\partial }{\partial \tau}\trace(\Pi R_{\mathrm{d}}(\tau)^{-1})\right],
\end{align}
where $\mu_k>0$ denotes the step size and $P_{\mathfrak{T}} [\cdot]$ denotes projection into the set $\mathfrak{T}$. For large $k$, we can replace $R_{\mathrm{d}}(\tau)$ with $R_{\mathrm{c}}(\tau)$ while noting that 
\begin{align*}
\bigg[\frac{\partial R_{\mathrm{c}}(\tau)}{\partial \tau_i}\bigg]_{ji}&=\bigg[\frac{\partial R_{\mathrm{c}}(\tau)}{\partial \tau_i}\bigg]_{ij}\\
&=\begin{cases}
-\displaystyle2\int_{-\infty}^{\infty} f''_i(t-\tau_i)f'_i(t-\tau_i)\mathrm{d}t, & i=j,\\
-\displaystyle\int_{-\infty}^{\infty} f''_i(t-\tau_i)f'_j(t-\tau_j)\mathrm{d}t, & i\neq j,
\end{cases}
\end{align*}
and, similarly, 
\begin{align*}
\bigg[\frac{\partial R_{\mathrm{c}}(\tau)}{\partial \tau_i}\bigg]_{q\ell}=0,\quad \forall q,\ell\neq i.
\end{align*}

%
%
%
%
%
%

\section{Numerical Example}
In this section, we use an illustrative example to demonstrate the results of this paper. For this purpose, let us consider three generic loads depicted in Figure~\ref{fig:example}. These loads possess simple forms for ease of demonstration. Assume that the additive noise is Gaussian with zero mean and standard deviation $\sigma_w=0.1$.

\begin{figure}
\begin{tabular}{c}
\begin{tikzpicture}
\node[] at (0,0) {\includegraphics[width=1\linewidth]{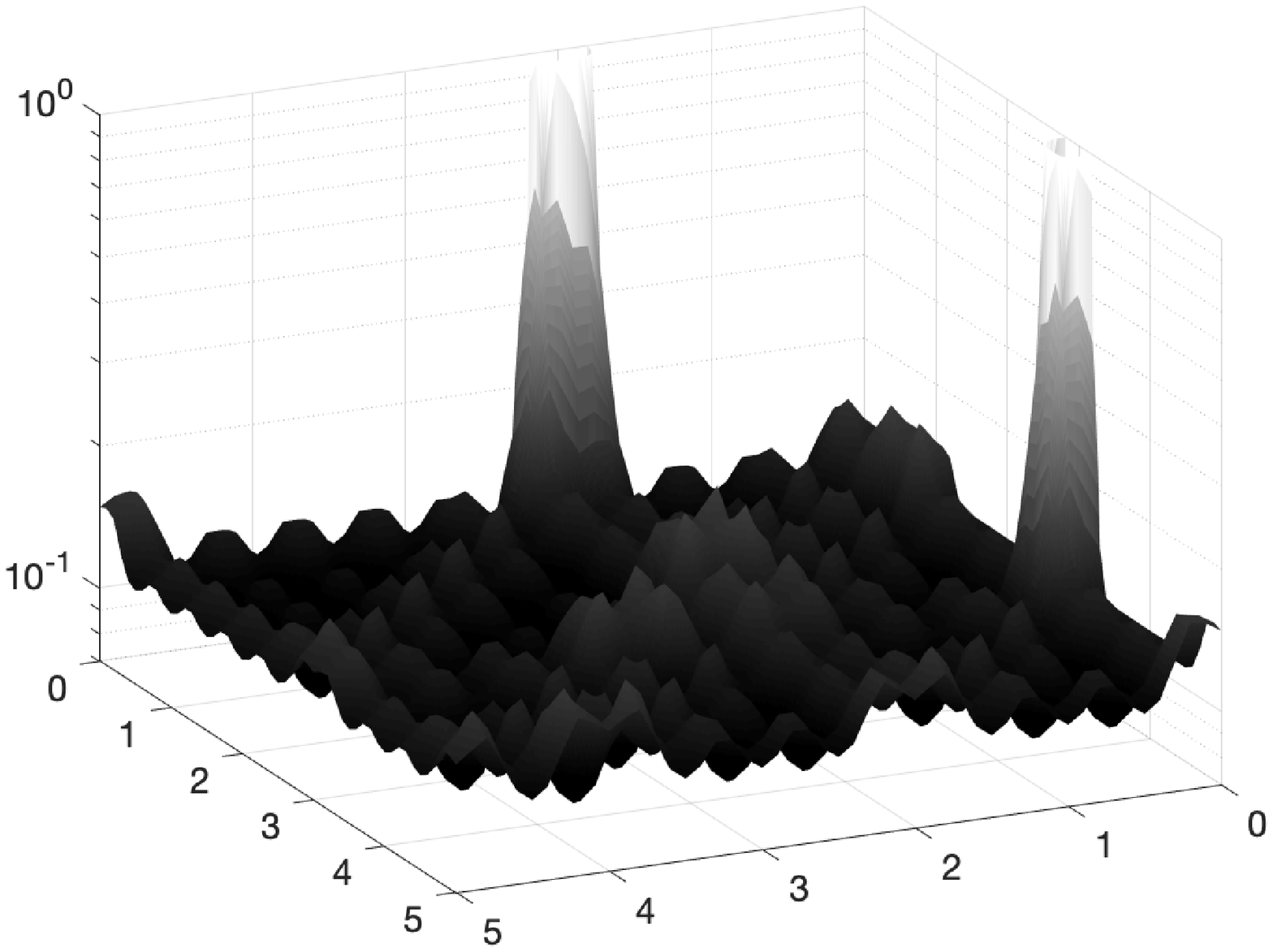}};
\node[] at (+1.7,-2.8) {$\tau_2$};
\node[] at (-2.6,-2.3) {$\tau_1$};
\node[rotate=90] at (-4.2,.5) {$\trace(R_{\mathrm{d}}(\tau)^{-1})$};
\end{tikzpicture}
\\[-2em]
\begin{tikzpicture}
\node[] at (0,0) {\includegraphics[width=1\linewidth]{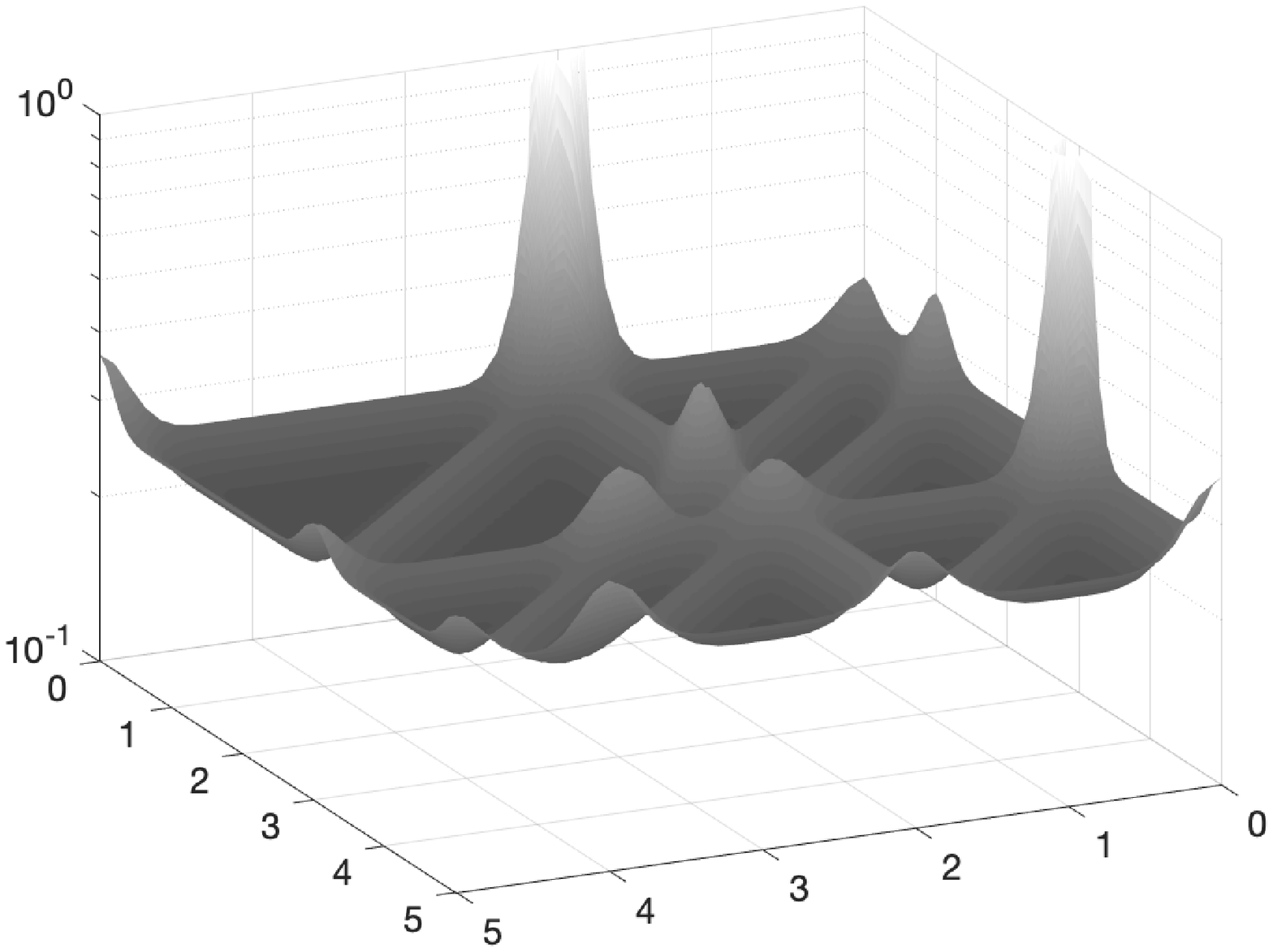}};
\node[] at (+1.7,-2.8) {$\tau_2$};
\node[] at (-2.6,-2.3) {$\tau_1$};
\node[rotate=90] at (-4.2,0.5) {$\trace(R_{\mathrm{c}}(\tau)^{-1})$};
\end{tikzpicture}
\end{tabular}
\vspace{-.3in}
\caption{
\label{fig:performance_bound}
Lower bound on the estimation error of any non-intrusive load monitoring algorithm in Theorem~\ref{tho:1} [top] and Corollary~\ref{cor:2} [bottom] versus scheduling delays of the first two loads.
}
\end{figure}

Assume that the non-intrusive load monitoring algorithm has access to the total measurements at $\{0,0.5,\dots,9.5,10\}$. Figure~\ref{fig:performance_bound} [top] illustrates the lower bound on the estimation error of any non-intrusive load monitoring algorithm in Theorem~\ref{tho:1} versus scheduling delays of the first two loads $\tau_1,\tau_2$ when fixing $\tau_3=0$. Figure~\ref{fig:performance_bound} [bottom] illustrates the lower bound on the estimation error of any non-intrusive load monitoring algorithm in Corollary~\ref{cor:2} versus scheduling delays of the first two loads. Clearly, these two bounds are very close to each other due to the high number of measurements in the discrete case.

Furthermore, as expected, the lower bounds in Theorem~\ref{tho:1} and Corollary~\ref{cor:2} are non-convex in the decision variables of the scheduling problem $\tau$ and possess many local optima. However, there are two points for which the lower bounds become very large, in fact these points are the globally-optimal privacy-preserving schedules when fixing $\tau_3=0$. They corresponds to schedules for which its is impossible to determine if the second load is scheduled first or the third load. Therefore, the estimation error of any non-intrusive load monitoring algorithm is large. Therefore, these are the most privacy-preserving schedules. 

\begin{figure}
	\begin{tikzpicture}
	\node[] at (0,0) {\includegraphics[width=1\linewidth]{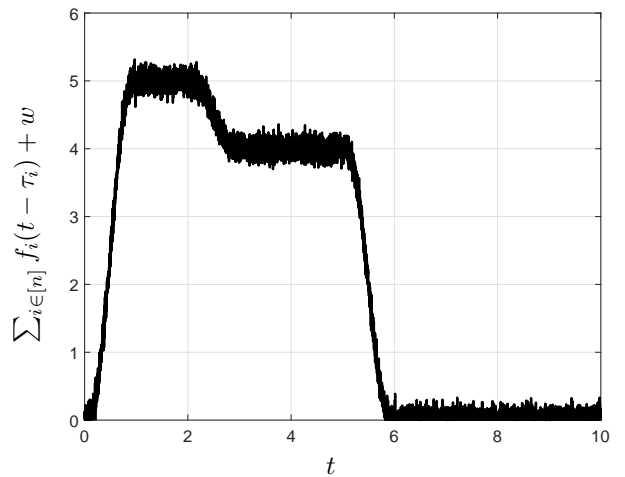}};
	\node[] at (0,-3.2) {$t$};
	\node[rotate=90] at (-4,0) {$\sum_{i\in\range{n}}f_i(t-\tau_i)+w$};
	\end{tikzpicture}
	\caption{Total noisy consumption for the privacy-preserving schedule.}	
\end{figure}

\section{Conclusions and Future Research}

We proved that the expected estimation error of non-intrusive load monitoring algorithms is lower bounded by the trace of the inverse of the cross-correlation matrix between the derivatives of the load profiles of the appliances. We developed privacy-preserving load scheduling policies by maximizing the lower bound on the expected estimation error of non-intrusive load monitoring algorithms. Future work can focus on experimental demonstration and validations of these results with off-the-shelf non-intrusive load monitoring algorithms.

\bibliographystyle{ieeetr}
\bibliography{ifacconf}     

\end{document}